\documentclass[onecolumn,preprintnumbers,amsmath,amssymb,nofootinbib]{revtex4}
\usepackage{graphicx,color}
\usepackage{amsmath,amssymb,latexsym}
\usepackage{bbm}
\usepackage{psfrag}

\begin{document}

\title{Ensemble properties of high frequency data and intraday trading rules}

\author{Fulvio Baldovin}
\affiliation{
Dipartimento di Fisica e Astronomia, Sezione INFN, CNISM, 
Universit\`a di Padova,
 Via Marzolo 8, I-35131 Padova, Italy
}

\author{Francesco Camana}
\affiliation{
Dipartimento di Fisica e Astronomia, Sezione INFN, CNISM, 
Universit\`a di Padova,
 Via Marzolo 8, I-35131 Padova, Italy
}

\author{Massimiliano Caporin}
\affiliation{
Dipartimento di Scienze Economiche ed Aziendali, Universit\`a
  di Padova, Via del Santo, I-35123 Padova,
  Italy}
  
\author{Michele Caraglio}
\affiliation{
Dipartimento di Fisica e Astronomia, Sezione INFN, CNISM, 
Universit\`a di Padova,
 Via Marzolo 8, I-35131 Padova, Italy
}

\author{Attilio L. Stella}
\affiliation{
Dipartimento di Fisica e Astronomia, Sezione INFN, CNISM, 
Universit\`a di Padova,
 Via Marzolo 8, I-35131 Padova, Italy
}




\begin{abstract}
Regarding the intraday sequence of high frequency returns of the S\&P index 
as daily realizations of a given stochastic process, 
we first demonstrate that the scaling properties of the 
aggregated return distribution  
can be employed to define a martingale stochastic model which consistently replicates
conditioned expectations of the S\&P 500 high frequency data in the morning of
each trading day. Then, a more general formulation of the above scaling properties
allows to extend the model to the afternoon trading session.
We finally outline an application in which conditioned forecasting is used
to implement a trend-following trading strategy capable of  
exploiting linear correlations present in the S\&P dataset and absent in the model.
Trading signals are model-based and not derived from chartist criteria.
In-sample and out-of-sample tests indicate that the model-based trading strategy performs
better than a benchmark one established on an asymmetric GARCH process, and 
show the existence of small arbitrage opportunities.
We remark that in the absence of linear correlations the trading profit would vanish and discuss 
why the trading strategy is
potentially interesting to hedge volatility risk for S\&P index-based 
products.
\\
\medskip
\textit{keywords:} Anomalous scaling; Memory; Intraday returns; Intraday strategy.

\end{abstract}

\maketitle


\section{Introduction}\label{S1}
Recent studies of high frequency (HF) data for foreign exchange (FX) rates
\cite{bMCg,bbcs,seemann} regarded the many daily realizations of asset
returns as constituting a statistical ensemble of histories
\footnote{Where a \textit{statistical ensemble} is a collection of
  elements, such as a collection of realizations of a stochastic
  process, with given statistical properties.  In the cited
  contributions, each daily HF dataset constitutes a single element of
  the ensemble. Furthermore, each element of the ensemble is assumed
  to be a realization of the same underlying stochastic process. The
  properties of the process at a given time within the day 
  are estimated on the basis of ensemble
  statistics, i.e. by averaging over all available daily
  realizations.}. The single HF time series from which such ensembles
can be extracted are known to present clear periodic patterns, with
period of one day \citep[see for instance][]{admati,andersen_1,guillame,dacorogna,allez}.
For example, in the
EUR/USD case, the volatility over successive 10 minutes intervals
monotonically increases or decreases during specific time windows
within the day. 
This behaviour is observed by averaging the volatility on both a daily
and a weakly basis. The idea in \citet{bMCg,bbcs} is to endogenize the
periodic, nearly deterministic behavior of the EUR/USD exchange rate
HF volatility into a time inhomogeneous stochastic process for the
evolution of the asset. The analysis of the ensemble of daily
histories produced by such a process reveals also a peculiar form of
scaling obeyed by the probability density functions (PDF) of the
returns, once they are aggregated over intervals of variable span
within a fixed intraday window \cite[see also][]{muller}. 
We recall that the scaling symmetry is
a relation linking the marginal PDF of a stochastic process with the
duration of sampling or observation interval. In the simplest case of
time intervals starting at the beginning of the daily window, the
scaling property observed for EUR/USD exchange rate 
implies that the
PDF $p(r,\tau)$ for a return $r$ over an interval of width $\tau$,
once multiplied by a factor $\tau^D$, yields a specific function of
$r/\tau^D$:
\begin{equation}
p(r,\tau)= \frac{1}{\tau^D}\;\;g\left(\frac{r}{\tau^D}\right),
\label{eq_scaling}
\end{equation}
where $g$ is the scaling function, and $D$ is a suitable scaling
exponent. This form leads to $p(r/\lambda^D,\tau/{\lambda})= 
\lambda^{D} p(r,\tau)$, for arbitrary rescaling $\lambda$ of the
interval time-span $\tau$, and means that the PDF of the 
returns has a known structure, independently from the duration of
aggregation intervals. A process with such a property is thus called {\it
  self-similar}. 
Whenever the returns of a given asset satisfy Eq. (\ref{eq_scaling}), 
a (power-law) decrease with time $\tau$ in the volatility is associated to $D<1/2$, whereas
(power-law) volatility increases are related to $D>1/2$. 
However, self similarity contains more information than this. 
At a first level, it fixes the distribution and the behavior of
all the moments of the aggregated returns, 
and this opens for instance 
the possibility of unconditioned predictions 
associated to the quantile function of these distributions. 
More deeply, we will see in what follows that it also permits the construction of
a multivariate PDF for modeling the underlying process. This in turn enables
conditioned forecasting.

The HF scaling symmetry (\ref{eq_scaling}) addressed in
Refs. \cite{bMCg,bbcs} has specific interesting aspects: $g$ is non-Gaussian
and, for the first hours of the morning trading, 
$D$ is lower than $1/2$. On the basis of the stability
of the Gaussian density under time aggregation, one would expect
$D=1/2$ and a Gaussian $g$ for independent
returns. On the other
hand, sliding-window empirical analyisis of the returns distribution
of single historical time series reveals
a non-Gaussian scaling but with $D\simeq 1/2$. 
The deviation from Gaussianity  
is an anomalous scaling feature
generally ascribed to long range dependence (as revealed, e.g., by the
presence of volatility clustering and heteroskedasticity with
persistent behaviours). However, for single time-series analysis of 
efficient markets historical data one still has
$D \simeq 1/2$ \cite{dacorogna,bs}. In the HF EUR/USD case mentioned above, besides
non-Gaussianity the strong deviation of $D$ from $1/2$ emphasizes the
anomalous character of the scaling and implies the time-inhomogeneity
(non-stationarity) of the returns' process.  
In particular, in \citet{bbcs} this time inhomogeneous scaling has been
attributed to a non-Markovian, strong dependence of the intraday
returns, as, e.g., manifested by the analysis of the
ensemble-averaged volatility autocorrelation function.  
This strong dependence amounts to an effective volatility clustering
phenomenon, partially hidden by the fact that the average $10$-minutes volatility
reduces in time when $D<1/2$. By imposing consistency with the
anomalous scaling of the aggregated return PDF, a martingale model 
has been then introduced for generating the histories
in the ensemble. Within this model, the PDF of each return retains
memory of the previous ones. Once properly calibrated, the model
replicates very well the non-linear statistical properties of the empirical HF
ensemble in the case of EUR/USD data. In particular, it is able to
reproduce the patterns observed for the autocorrelations
of absolute returns and squared returns obtained
from the HF data. 

In the present work we apply and extend the non-Markovian model introduced in
\citet{bbcs} to the description of the HF data of a different
financial asset, namely the S\&P 500 index. A first goal is to
show that the model is able to replicate the martingale morning features of an asset
different from FX exchange rates. 
We will then extend our approach to include the description of 
the first part of the afternoon trading, up
to 16:00 p.m., where a different, increasing, power-law behaviour of the volatility
can be associated to a more complicated scaling simmetry
than the one reported in Eq. (\ref{eq_scaling}). 

In order to outline a practical application of the above stochastic modeling, 
we will finally devise a trading strategy built on the unconditioned and 
conditioned quantile functions predicted by the model.
We will show that the proposed trading strategy exposes small arbitrage opportunities
related to linear correlations present in the S\&P dataset 
albeit by construction absent in our martingale model.
We also compute density forecasts 
using a simpler and more standard GARCH martingale process
\cite{engle,bollerslev}; namely, the asymmetric GJR model \cite{glosten}. 
The density forecasts obtained from our model and those of the GARCH
benchmark are used to define price bounds whose violation gives a
trading signal. The bounds might be interpreted as predicted supports
and resistances, or they could be read as a simulated price range with
a given confidence interval. Our trading approach is a trend following protocol
which thus belongs to the large
plethora of Technical Analysis-based  methods whose
performances have been studied by different authors
\cite[][among others]{lo,neely,park}. However, the indicators
used here to derive the trading signals are model-based and not derived
from a pure \textit{chartist} approach. In-sample and out-of-sample
tests demonstrate the existence of trading opportunities, leading,
anyhow, to relatively small average margins of profit. As a matter of
fact, those profits vary over time, reaching interesting values in
specific periods and always beating the average profits based on
the GARCH model forecasts.

The paper proceeds as follow. In subsection 1.1 we briefly describe
the data used within this study, while Section 2 is devoted to the
presentation of the model and to parameter calibration. 
Section 3 presents the version of the model suitable to describe a wider daily
window of market activity.
Section 4
deals with the construction of density forecasts, and Section 5
describes the trading strategy and reports the empirical
results. Section 6 concludes.

\subsection{Data extraction}\label{S1.1}
In this work we focus on 
the S\&P 500 index. We
consider a dataset ranging from September 30th  
1985 to October 19th 2010. After excluding those days for which the
records are not complete 
({\it e.g.}, for holidays or stock market anticipated closures),
the whole dataset includes $M=6179$ trading days.
Because of dataset limitations, 
we exclude the first and last half an hour of each daily trading session.
Such a choice has a further effect: the intra-daily periods with highest volatility,
i.e. opening and closing, are excluded from the analysis.
For each single day $l$ ($1\leq l\leq M$), in the first part
of our analysis, we extract the index values,
$s_t^{(l)}$, 
every 10 minutes between 10:00 a.m. (when we set $t=0$) and 13:20 a.m. ($t=20$), 
New York time. The reasons to choose a 10 minutes time interval will be made 
clearer in the following.

The empirical returns of the $l$-th day are thus defined 
as $r_t^{(l)}\equiv\ln s_t^{(l)}-\ln s_{t-1}^{(l)}$
and are regarded as specific realizations of stochastic variables $R_t$.
Our first task is thus to identify a proper analytical model for the
joint PDF of the returns $R_t$,
$p(r_1,r_2,\ldots,r_\tau)$, which correctly
reproduces the statistics of the ensemble 
$\left\{r^{(l)}_t\right\}_{\substack{t=1,\ldots,20\\l=1,\ldots,M}}$. 
Since we will assume a martingale stochastic modeling, we 
checked that linear correlation 
effects among consecutive returns are negligible to a reasonable
approximation\footnote{
The mean empirical linear correlation of 10-minute returns is
$ \dfrac{1}{20}  \sum_{t=1}^{20}  \dfrac{ \sum_{l}\;r_t^{(l)}\;r_{t+1}^{(l)}}
{\sqrt{ \sum_{l} \; \left(r_t^{(l)}\right)^2} \; \sqrt{
\sum_{l}\;\left(r_{t+1}^{(l)}\right)^2 }}
\simeq 0.05$.
}. 
In Section 3 we extend the analysis up to 16:00 p.m. New York time, 
our ensemble hence becomes 
$\left\{r^{(l)}_t\right\}_{\substack{t=1,\ldots,36\\l=1,\ldots,M}}$. 

\section{The model}\label{S2}
In \citet{bMCg,seemann} it has been suggested that HF financial
time-series for the FX exchange rates offer the opportunity to deal 
with many realizations of the same stochastic process every day.
The set of daily histories, restricted to suitable time-windows, can thus be 
regarded as an ensemble for testing the statistical properties of the 
underlying process. 
In fact these histories are not completely 
independent, because heteroskedasticity and volatility correlations, if estimated
by time averages along the whole time series $s_t$ from which the ensemble
$s_t^{(l)}$ is extracted, exceed the intraday range. 
However, for a large enough total number of days $M$ one can expect inter-day
correlations effects to compensate, allowing for a reliable statistics of
the postulated ensemble. 

The general ideas at the basis of the mathematical model 
adopted here to describe such an ensemble can be traced back in Refs. \cite{bs,bbcs}. 
Financial market returns in the HF ensemble relative to
the EUR/USD exchange rate are, to a first approximation, linearly
uncorrelated if taken over intervals of 10 minutes or more.
However, they are also dependent, as shown, e.g., by the nonzero volatility
autocorrelation function. As a simplifying assumption, we also disregard the
weak skewness of the returns distribution,  
whose inclusion in the modeling framework is left to a future development.
Based on these evidences and on the existence of 
time inhomogeneous anomalous scaling properties (see below), 
it has been proposed that the joint PDF's of these returns have the 
form of convex combinations of the joint PDF's of processes with 
independent Gaussian increments of different, time dependent widths.
This implies that the scaling function $g$ of the aggregated return PDF
is also a convex combination of Gaussians of variable width \cite{stella,peirano}, as often 
assumed in phenomenological studies of anomalous scaling 
\cite[see, e.g.,][]{bp}.

The starting point of our modeling is the assumption of the
validity of the scaling symmetry, Eq. (\ref{eq_scaling}),
for the aggregated return over the time scale $\tau$,
$\sum_{t=1}^\tau R_t$, with a non-Gaussian scaling function $g$ and $D<1/2$.
This assumption is reasonably well verified in the morning window ($ 1 \le \tau \le 20$).
Consistently with a power-law decay of the volatility
$\sqrt{\mathbb E[R_t^2]}$ during the morning trading hours,
an exact scaling symmetry with $D<1/2$ also implies a power-law behavior for all
the existing moments of the distribution, according to
\begin{equation}
\label{eq_moments}
\mathbb E\left[\left|R_1+\cdots+R_\tau\right|^q\right]
=\mathbb E\left[\left|R_1\right|^q\right]
\;\tau^{q\,D}.
\end{equation}

The main feature of the time-inhomogeneous model described in
\citet{bbcs} is the construction of the multivariate returns PDF,
$p(r_1,r_2,\dots,r_\tau)$,~\footnote{
For simplifying our notations, we remove the stochastic variables subscripts
to the PDF's symbols. Explicit inclusion of the arguments thus discriminates
whether for instace we are talking about a single-point marginal PDF, or about 
a many-point joint PDF. 
}
on the basis of the scaling symmetry
valid for the marginal PDF of the aggregated returns.
Indeed, the joint PDF for the returns is    
reconstructed as
\begin{equation}
{ p(r_1,r_2, \dots, r_\tau)=\int_0^\infty d\sigma \rho(\sigma)}
\prod_{t=1}^{\tau}\frac{\exp\left(-{{r_t^2}\over{2\sigma^2
      a_t^2}}\right)}{\sqrt{2\pi\sigma^2 a_t^2}}, 
\label{eq_joint_probability}
\end{equation}
where
\begin{equation}
\label{eq_a_t}
a_t\equiv\left[t^{2D}-(t-1)^{2D}\right]^{1/2},
\end{equation}
$\rho(\sigma)\geq0$ is a PDF for a mixture of Gaussian processes
with different widths $\sigma$
\cite[see, e.g.,][]{clark,wx}, and the scaling function $g$ is
given by 
\begin{equation}
\label{eq_g}
{g(r)=\int_0^\infty \rho(\sigma)} {e^{-{{r^2}\over{2\sigma^2}}}\over{\sqrt{2\pi\sigma^2}}}d\sigma. 
\end{equation}
The coefficients $a_t$ determine the
time-inhomogeneity of the variables $R_t$, which is also manifest in a peculiar
scaling form for its marginal PDF, $p(r_t)$. Namely,
\begin{equation}
\label{eq_scaling_single_return}
p(r_t)=\dfrac{1}{a_t}\;g\left(\dfrac{r_t}{a_t}\right).
\end{equation}
Only for $D=1/2$ the variables $R_t$  become
identically distributed.  See Refs. \cite{bs,stella} for additional details on 
the derivation of the joint PDF.

The next step is the identification of a proper parametrization for
the scaling function $g$.  
As $\sigma$ is a measure of volatility, many possible modelings are
available in the literature \cite[see, $e.g.$,][]{micci,bp,peirano}.
A convenient way of 
representing fat-tailed scaling
functions as those revealed by empirical analyses in finance, with the additional 
benefit that the integration over $\sigma$ can be performed explicitly,
is by using an 
{\it inverse-gamma} density for $\sigma^2$. 
With this particular choice $\rho(\sigma)$ becomes 
\begin{equation}
\label{eq_inverse_gamma}
\rho(\sigma)=\frac{2^{1-\frac{\alpha}{2}}}{\Gamma(\frac{\alpha}{2})}
\;\frac{\beta^\alpha}{\sigma^{\alpha+1}}
\text{e}^{-\frac{\beta^2}{2\sigma^2}},
\end{equation}
where $\alpha>0$ and $\beta>0$ are a {\it form} and a {\it scale} parameter, respectively,
and $\Gamma$ is the Euler's gamma function.
The resulting scaling function is then a Student's t-distribution,
\begin{equation}
\label{eq_scaling_function}
g(r)=\frac{\Gamma(\frac{\alpha+1}{2})}{\sqrt{\pi}\;\Gamma(\frac{\alpha}{2})}
\;\dfrac{1}{\beta}
\;\left(1+\dfrac{r^2}{\beta^2}\right)^{-\frac{\alpha+1}{2}}.
\end{equation}
This $g(r)$ has a power-law decay for large $|r|$ with exponent $\alpha+1$,
whereas $\beta$ simply sets the scale of its width.
Moreover, with the choice in 
Eq. (\ref{eq_inverse_gamma}) for $\rho(\sigma)$, we have
\begin{equation}
\label{eq_moment_1}
\mathbb E[|R_1+\cdots+R_\tau|^q]
=\dfrac{
\Gamma\left(
\dfrac{q+1}{2}
\right)
\;\Gamma\left(
\dfrac{\alpha-q}{2}
\right)
\;\beta^q}
{
\sqrt{\pi}
\;\Gamma\left(
\dfrac{\alpha}{2}
\right)
}
\;\tau^{q\,D}.
\end{equation}

In this paper we discuss the application of this model to the S\&P 500
index, working with  
a larger set of realizations ($M=6179$) with respect to the EUR/USD
dataset used in \citet{bbcs}. 
We find that
the general features described in Ref. \cite{bbcs} 
also characterize the S\&P 500 index
within the morning time window, and that the model described above
reproduces well these features.
In both cases the volatility at $10$ minutes intervals tends to decrease
in the chosen window.
In the Appendix we report additional elements about
the small linear correlations of returns at the
$10$ minutes time-scale and the simultaneous presence
of strong non-linear correlations. 
In the next Section we report instead the empirical evidence used to
calibrate the three parameters ($D,\alpha$, and $\beta$) for the morning
trading session.

\subsection{Morning calibration}
The scaling exponent $D$ and the scaling function $g$ play here a central 
role as they determine respectively the $a_i$'s and $\rho(\sigma)$
appearing in  
the joint PDF [Eq. (\ref{eq_joint_probability})] 
of a given daily realization of the process.
We adopt a multi-step calibration procedure. 
At first, we calibrate $D$, and then we use $D$ in order to
obtain a data-collapse which allows identifying $\alpha$ and $\beta$
[thus $\rho(\sigma)$ and $g$ -- see Eqs. (\ref{eq_inverse_gamma}) and (\ref{eq_g})].

\begin{figure}
\begin{center}
\resizebox*{8cm}{!}{\includegraphics{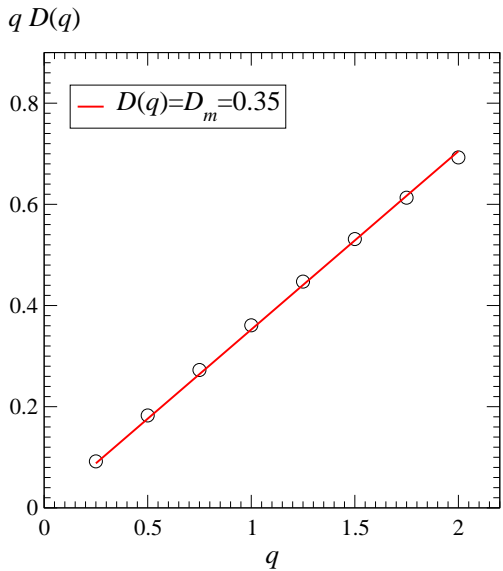}}
\caption{Scaling behaviour of the non-linear moments.}
\label{fig:nl_moments}
\end{center}
\end{figure}

A quantitative way to calibrate $D$ is offered by the analysis of the
moments of \mbox{$R_1+\cdots+R_\tau$},
since in the presence of a simple scaling symmetry the moments 
of the aggregated returns satisfy  Eq. (\ref{eq_moments}), where 
only $\mathbb E[|R_1|^q]$ dependens on $g$.
The logarithm of Eq. (\ref{eq_moments}) gives thus a linear relation vs $q$, with slope
equal to $D$.
Calling $q\,D(q)$ the exponent empirically
estimated for the power law in Eq. (\ref{eq_moments}),
in Fig.~\ref{fig:nl_moments} we plot the results of the linear
regression, $q\,D(q)\simeq q\,D_m$, of the logarithm of
the empirical moments as a function of $\tau$
\begin{equation}
\overline{|r_1+\cdots+r_\tau|^q} \equiv
{1\over M} \sum_{l=1}^{M} |r_1^{(l)}+\cdots+r_\tau^{(l)}|^q\quad (0<q<2).
\end{equation} 
The resulting regression slope $D(q)=D_m\simeq 0.35$, 
identifies the value of $D$ for the morning trading session 
and is consistent with the 
anomalous
scaling symmetry [Eq. (\ref{eq_scaling})] 
satisfied by the empirical S\&P 500 data. This value
is very close to that estimated for FX exchange rates 
in Refs. \cite{bMCg,bbcs,seemann}.
This is probably due to the fact that the $10$-minutes volatility
has a very similar decreasing trend in the window considered in the two cases.
It is important to point out that we limit our scaling analysis to
$q\lesssim 2$ because of an evident multiscaling behaviour for
$q\gtrsim 2$ \cite[see also][]{wang}.

\begin{figure}
\begin{center}
\resizebox*{8cm}{!}{\includegraphics{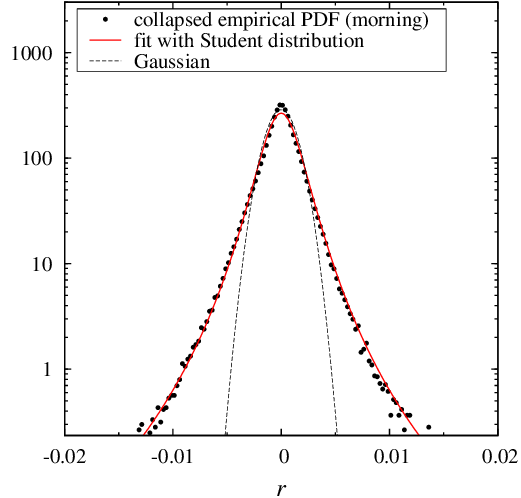}}
\caption{The scaling function for the S\&P after collapse of aggregated and marginal returns 
  (points) in the morning time window. 
  The red line is a fit to the points with the Student distribution given in Eq. (\ref{eq_scaling_function})
.}
\label{fig:scaling_fcn}
\end{center}
\end{figure}

Taking advantage of the knowledge of the (morning) scaling exponent $D_m=0.35$, 
we now fix a relation between $\alpha$ and $\beta$, still analyzing the 
moments of the aggregated returns. 
Thus, a least squares fitting of $\overline{|r_1+\cdots+r_\tau|}$ using 
Eqs. (\ref{eq_moments}) with $q=1$ and $D=D_m$ 
fixes $\mathbb E[|R_1|]\simeq1.5\cdot10^{-3}$.
Through Eq. (\ref{eq_moment_1}), this gives, e.g., $\beta$ as a function of $\alpha$.

Finally, we data-collapse the empirical PDF of the aggregated returns 
$R_1+\cdots+R_\tau$ and of the marginal returns $R_t$,
using Eq. (\ref{eq_scaling}) with $\tau=1,2,\ldots,20$ 
and Eq. (\ref{eq_scaling_single_return}) with $t=1,2,\ldots,20$, respectively
(see Fig.~\ref{fig:scaling_fcn}). 
A least squares optimization of this data-collapse with
Eq. (\ref{eq_scaling_function}) and $\beta$ fixed as described above
allow us to determine $\alpha\simeq3.29$.

In summary the (in-sample) model calibration for the morning trading session 
gives \mbox{$D=D_m=0.35$}, $\alpha=3.29$, \mbox{$\beta=\beta_m=2.5\cdot10^{-3}$}. 
These notations anticipate that in the extension of the model to the afternoon session
we keep the form parameter $\alpha$ fixed, 
and change the value of the scaling exponent $D$ and of the scale parameter $\beta$.
As explained below, 
this is done consistently with the empirical analysis of the afternoon dataset.

\section{Extension of the model to the afternoon trading session}
For the purpose of extending our modeling, let us first recast 
Eqs. (\ref{eq_joint_probability}) and (\ref{eq_inverse_gamma})
by performing the transformation $\sigma\mapsto\sigma/\beta$:
\begin{equation}
{ p(r_1,r_2, \dots, r_\tau)=\int_0^\infty d\sigma \rho^\prime(\sigma)}
\prod_{t=1}^{\tau}\frac{\exp\left(-{{r_t^2}\over{2\sigma^2
      a_t^2\beta^2}}\right)}{\sqrt{2\pi \sigma^2 a_t^2\beta^2}} \; , 
\label{eq_joint_probability_recasted}
\end{equation}
\begin{equation}
\label{eq_inverse_gamma_recasted}
\rho^\prime(\sigma)=\frac{2^{1-\frac{\alpha}{2}}}{\Gamma(\frac{\alpha}{2})}
\;\frac{1}{\sigma^{\alpha+1}}
\text{e}^{-\frac{1}{2\sigma^2}} \; .
\end{equation}
Within this formulation it is clearer the role of $\beta$  
as a scale parameter of the returns variables.

\begin{figure}
\begin{center}
\resizebox*{8cm}{!}{\includegraphics{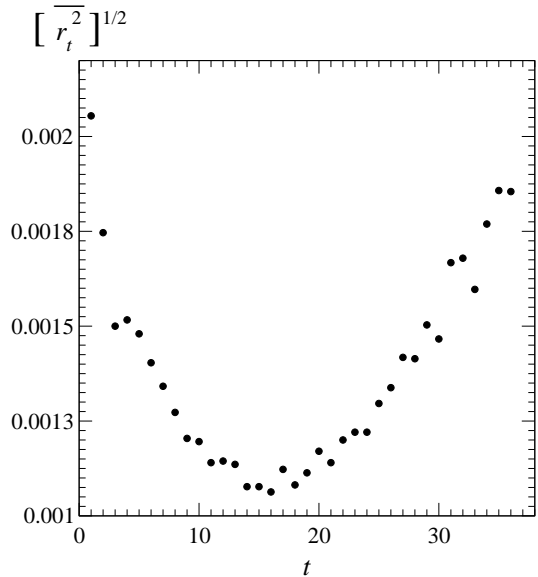}}
\caption{Empirical volatility of the S\&P dataset, during a whole trading day.}
\label{fig:sep_volatility}
\end{center}
\end{figure}

Fig. \ref{fig:sep_volatility} shows that, consistently with our modeling for the 
morning session, during the
morning hours the empirical return volatility 
$\sqrt{\overline{r_t^2}}$ decreases.
However, around $t=t^\ast\simeq20$ (13:20 New York time)
this trend is definitely inverted.
Within our approach, an unconditioned (power-law) volatility increase can only 
be associated with a scaling exponent $D>1/2$.
The simplest way to extend our model is thus to keep the validity of
Eq. (\ref{eq_inverse_gamma_recasted}) and generalize
Eqs. (\ref{eq_a_t}) and (\ref{eq_joint_probability_recasted})
by introducing a time dependence 
of the scaling exponent $D\mapsto D_t$ and of the scale parameter 
$\beta\mapsto\beta_t$:
\begin{equation}
\label{eq_a_t_extended}
a_t\equiv\left[t^{2D_t}-(t-1)^{2D_t}\right]^{1/2} \; ,
\end{equation}
\begin{equation}
{ p(r_1,r_2, \dots, r_\tau)=\int_0^\infty d\sigma \rho^\prime(\sigma)}
\prod_{t=1}^{\tau}\frac{\exp\left(-{{r_t^2}\over{2\sigma^2
      a_t^2\beta_t^2}}\right)}{\sqrt{2\pi \sigma^2 a_t^2\beta_t^2}} \; . 
\label{eq_joint_probability_extended}
\end{equation}
Specifically, a natural choice which leaves unchanged all the morning features and
introduces the aforementioned afternoon volatility power law increase is
\begin{equation}
D_t\equiv\begin{cases}
D_m & \mbox{if } 1\leq t\leq t^\ast \; ;\\
D_a & \mbox{if } t^\ast \leq t\leq 36 \; , 
\end{cases}
\end{equation}
\begin{equation}
\beta_t\equiv\begin{cases}
\beta_m & \mbox{if } 1\leq t\leq t^\ast \; ;\\
\beta_a & \mbox{if } t^\ast \leq t\leq 36 \;, 
\end{cases}
\end{equation}
with $D_a>1/2$.
Explicit integration over $\sigma$ leads to
\begin{equation}
\label{eq_joint_explicit}
p (r_1,\ldots, r_\tau) 
= \left( \prod_{t=1}^\tau \dfrac{1}{a_t\,\beta_t } \right) 
\;\dfrac{\Gamma \left( \frac{\alpha+\tau}{2} \right) }{\pi^{\frac{\tau}{2}} 
\;\Gamma \left( \frac{\alpha}{2} \right)} 
\;\left[  1 + \left( \dfrac{r_1}{a_1\,\beta_1}\right)^2 
+\cdots+ \left( \dfrac{r_\tau}{a_\tau\,\beta_\tau} \right)^2 \right] ^{-\frac{\alpha + \tau}{2}}.
\end{equation}
In the Appendix we show that this scaling-inspired, martingale 
extension of the model well reproduces
the non-linear correlation structure of empirical returns, both during the morning 
and the afternoon trading sessions. 

\begin{figure}
\begin{center}
\resizebox*{8cm}{!}{\includegraphics{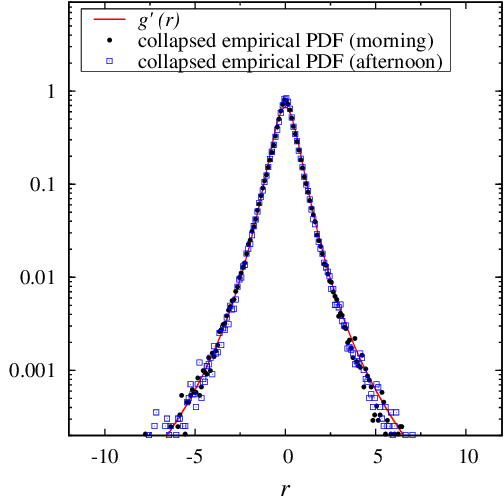}}
\caption{The scaling function for the S\&P after collapse of aggregated and marginal returns in the morning and afternoon time windows. Red line is the Student distribution given in Eq. (\ref{eq_scaling_function_extended})
.}
\label{fig:scaling_fcn_allday}
\end{center}
\end{figure}

With this extension of the model, the scaling simmetry takes the form:
\begin{equation}
p(r,\tau)= \frac{1}
{\lambda(\tau,t^\ast)
}
\;\;g'\left(\frac{r}{\lambda(\tau,t^\ast)}\right),
\label{eq_scaling_extended}
\end{equation}
with
\begin{equation}
\label{eq_lambda}
\lambda(\tau,t^\ast)
\equiv
\;\left(\sum_{t=1}^\tau a_t^2\,\beta_t^2\right)^{1/2}
=\;\begin{cases}
\beta_m\,\tau^{D_m}
& \mbox{if } 1\leq \tau\leq t^\ast;\\
\left[
\beta_m^2\,(t^\ast)^{2D_m}+\beta_a^2\left[\tau^{2D_a}-(t^\ast)^{2D_a}\right]
\right]^{1/2} 
& \mbox{if } t^\ast \leq \tau\leq 36, 
\end{cases}.
\end{equation}
and
\begin{equation}
\label{eq_scaling_function_extended}
g'(r)=\frac{\Gamma(\frac{\alpha+1}{2})}{\sqrt{\pi}\;\Gamma(\frac{\alpha}{2})}
\;\left(1+r^2\right)^{-\frac{\alpha+1}{2}}.
\end{equation}
Clearly for $\tau \le t^\ast$, Eq. (\ref{eq_scaling_extended}) reduces to
Eq. (\ref{eq_scaling}), taking into account Eq. (\ref{eq_lambda}) and the
fact that $g(x)=\frac{1}{\beta_m} g' \left( \frac{x}{\beta_m} \right) $.
In addition, the marginal single-return distribution preserves a scaling form which 
generalizes Eq. (\ref{eq_scaling_single_return}) into 
\begin{equation}
\label{eq_scaling_single_return_extended}
p(r_t)=\dfrac{1}{a_t\,\beta_t}\;g'\left(\dfrac{r_t}{a_t\,\beta_t}\right)
\end{equation}
with $g'(r)$ given by Eq. (\ref{eq_scaling_function_extended}).
Fig. \ref{fig:scaling_fcn_allday} shows the consistency  of these
extended scaling laws with the empirical evidence.
Finally, Eq. (\ref{eq_moment_1}) for the 
moments of the aggregated returns distribution 
is now replaced by 
\begin{equation}
\mathbb E\left[\left|R_1+\cdots+R_\tau\right|^q\right]
=
\dfrac{
\Gamma\left(
\dfrac{q+1}{2}
\right)
\;\Gamma\left(
\dfrac{\alpha-q}{2}
\right)
}
{
\sqrt{\pi}
\;\Gamma\left(
\dfrac{\alpha}{2}
\right)
}
\;\left[\lambda(\tau,t^\ast)\right]^{q}.
\label{eq_moments_extended}
\end{equation}

\subsection{Afternoon calibration and out-of-sample calibration}
While $D_m$, $\alpha$, $\beta_m$ have been (in-sample) calibrated as
described in Section 2.1, $D_a$ and $\beta_a$ remain to be identified.

\begin{figure}
\begin{center}
\resizebox*{8cm}{!}{\includegraphics{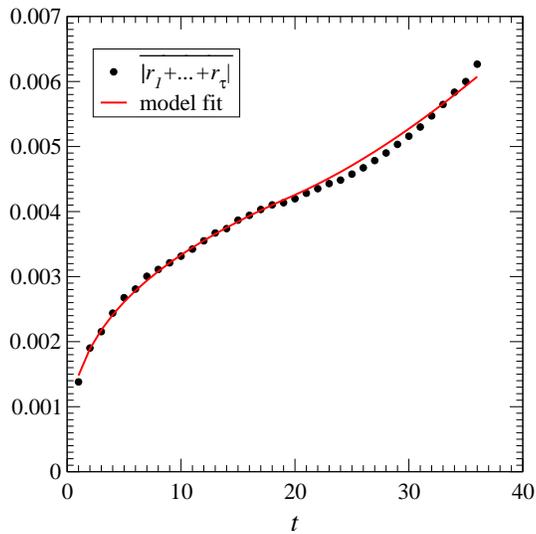}}
\caption{Empirical first moment of the aggregated returns during a whole trading day and 
data fit using the model.}
\label{fig:sep_2nd_aggregated}
\end{center}
\end{figure}

We fix $\beta_a$ by imposing  a matching condition on the width of 
the marginal PDF $p(r_{t^\ast})$. Namely, 
\begin{equation}
\beta_a=
\beta_m
\;\dfrac{
\left[(t^\ast)^{2D_m}-(t^\ast-1)^{2D_m}\right]^{1/2}
}{
\left[(t^\ast)^{2D_a}-(t^\ast-1)^{2D_a}\right]^{1/2}
}
\end{equation}
In this way, the only parameter which remains to calibrate is $D_a$. 
Again we opt for 
a least squares fitting of $\overline{|r_1+\cdots+r_\tau|}$ for $\tau>t^\ast$, 
using Eqs. (\ref{eq_moments_extended}) with $q=1$.
The result is
$D_a\simeq1.31$ and thus $\beta_a\simeq7.5\cdot10^{-5}$.
Fig. \ref{fig:sep_2nd_aggregated} 
displays that indeed our model well reproduces the empirical 
first moment of the aggregated returns, both in the morning and in the afternoon
trading sessions. 
Summarizing our results for the in-sample calibration, we have thus
\mbox{$(D_m,D_a,\alpha,\beta_m,\beta_a)=(0.35,1.31,3.29,2.5\cdot10^{-3},7.5\cdot10^{-5})$}.

\begin{figure}
\begin{center}
\resizebox*{8cm}{!}{\includegraphics{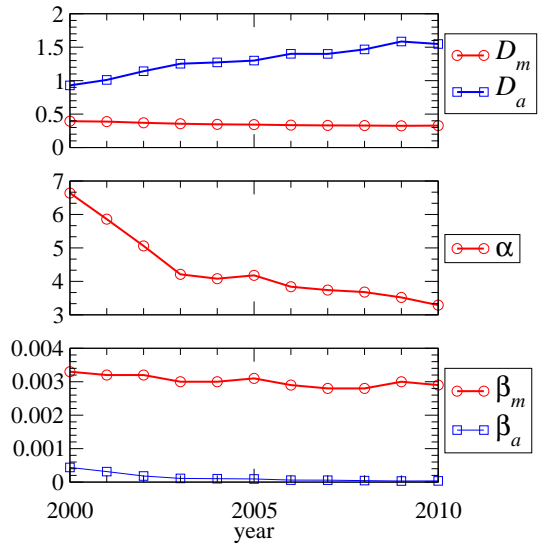}}
\caption{Fitting parameters for the out-of-sample analysis (using 15
  previous years).} 
\label{fig:parameters}
\end{center}
\end{figure}

However, we are interested in performing out-of-sample analyses
also.  In such cases, starting from our 25-years database, we decided
to use the first 15 years of data (from 1985 to 1999) to identify specific
values for the parameters $(D_m,D_a,\alpha,\beta_m,\beta_a)$ 
to be used to realize the trading strategy for the year
2000. We then repeatedly shifted year by year, always using the 15
previous years to calibrate the model, until 2010.
So, $e.g.$, we used the data from 1994 to 2008 to fit the
scaling function for the strategy to be used for year 2009.
Fig.~\ref{fig:parameters} shows the optimal parameter values 
used for the implementation of the
out-of-sample strategies from 2000 to 2010. We observe that
$D_a$ is almost stable in the out-of-sample evaluation while
$D_m$ tends to increase with time. Furthermore, the scale paramenters
$\beta_m$ and $\beta_a$ are quite stable while $\alpha$ decreases with time.

\section{Density forecasts and trading signals}
The model proposed here is a martingale with implicit 
forecasting capabilities of the asset's fluctuations. 
Assuming correct model specifications and
given the parameter values and the first $t_p$ daily returns, 
the model determines, under the martingale assumption, the conditional PDF
of the returns at $t>t_p$. 
Of course, this also gives 
the conditional PDF of any aggregation of
subsequent returns within the validity of the extended model. 
Using these density forecasts it is possible to construct a trading
strategy.
We consider in the following two different trading
approaches: the first makes use of the opening value on each day $l$ of the traded
asset at the beginning of the chosen time-window, $s_0^{(l)}$, 
and then bases  the density forecasts only on this value. We call it
\textit{unconditioned trading} since no information
coming from the daily returns is exploited ($t_p=0$). 
In the second approach, besides $s_0^{(l)}$, we use the information
contained in the first daily returns (up to $t_p>0$) 
for the density forecasts of the remaining part of the daily window.

With both trading approaches it is possible to extract trading signals
from density forecasts: given a certain value $0< \mathit{Q} <1/2$,
if at a certain time within the intra-daily
range under study the observed market price is above (below) the 
quantile function in $1-\mathit{Q}$ ($\mathit{Q}$) 
of the predicted price PDF we have a buy (sell)
signal.
These signals are regarded as potential warnings of the presence of a trend, which, 
altough absent in our stochastic modeling, affects the real assets' dynamics, as 
testified by the existence of a small, non-vanishing empirical linear correlations.  
Indeed, the trend-following trading strategy outlined below gives close to zero profit if
applied on histories generated numerically on the basis of our martingale model.

\begin{figure}
\begin{center}
\resizebox*{10cm}{!}{\includegraphics{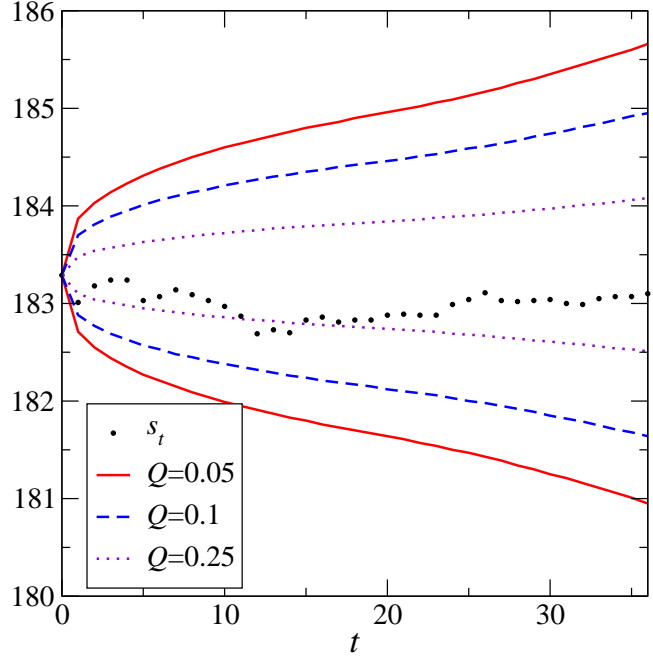}}
\caption{Upper and lower expected index values for October 4th 1985,
confronted with real prices (black circles).
Lines are linear piece-wise interpolations.} 
\label{fig:day_1}
\end{center}
\end{figure}

\subsection{Unconditioned trading signals}\label{S2.1.1}
In this case, 
for each day $l$, we calculate the quantile function for
the expected price distribution at time $t$,
conditioned to the opening value $s_0^{(l)}$ only. 
Note that our approach, in computing the density forecasts,
does not rely on specific previous-days information on the asset.
It is assumed that all the information contained in the past
observations it is contained in the model parameters, namely 
\mbox{$(D_m,D_a,\alpha,\beta_m,\beta_a)$}.  
According to Eqs. (\ref{eq_scaling_extended})
and (\ref{eq_scaling_function_extended}),
the PDF for the aggregated return
$R_1+\cdots+R_\tau$ is 
given by: 
\begin{equation}
p(r,\tau)
=\frac{\Gamma(\frac{\alpha+1}{2})}{\sqrt{\pi}\;\Gamma(\frac{\alpha}{2})}
\;\dfrac{1}{\lambda(\tau,t^\ast)}
\;\left[1+\left(\dfrac{r}{\lambda(\tau,t^\ast)}\right)^2\right]^{-\frac{\alpha+1}{2}},
\end{equation}
with $\lambda(\tau,t^\ast)$ as in Eq. (\ref{eq_lambda}).
The lower-bound values $r_{min,\tau}(\mathit{Q})$ of the expected aggregated returns
for the value $\mathit{Q}$ of the cumulative distribution 
are then obtained by numerically solving the equation
\begin{equation}
\mathit{Q}=\int_{-\infty}^{r_{min,\tau}(\mathit{Q})} dr\;p(r,\tau)
\end{equation}
with respect to $r_{min,\tau}(\mathit{Q})$.
Due to the parity of the
scaling function $g(r)$, the corresponding upper-bound values
$r_{max,\tau}(\mathit{Q})$ are simply obtained via sign flip:
$r_{max,\tau}(\mathit{Q})=-r_{min,\tau}(\mathit{Q})$. 

The lower  and upper expected
values, $s_{min,\tau}(\mathit{Q})$ 
and $s_{max,\tau}(\mathit{Q})$, respectively, 
of the asset price at time $\tau$ can then be easily calculated.
Indeed, given
$s_0^{(l)}$, the asset price $S_\tau$ is a
monotonic function of the $R_t$'s:
\begin{equation}
S_\tau=s_0^{(l)}\;{\rm e}^{\sum_{t=1}^\tau R_t}.
\end{equation}
Hence, the 
quantile function for $S_\tau$ is directly related to those of
$R_t$'s, once the daily opening value is given. 
Notice that in order to simplify our notations, we have dropped the 
dependence of $s_{min,\tau}(\mathit{Q})$ 
and $s_{max,\tau}(\mathit{Q})$ on the trading day.

Summarizing, for every choice of $\mathit{Q}$ with $0<\mathit{Q}<1/2$, for every time $t$ from 1
to 36 within each trading day, two price barriers are obtained, $s_{min,\tau}(\mathit{Q})$ 
and $s_{max,\tau}(\mathit{Q})$. According to our martingale modeling, with 
probability $1-2\mathit{Q}$ the price at time $t$ is placed between these
values.
For instance, in Fig.~\ref{fig:day_1}  the results of the in-sample analysis for
the day October 4th, 1985, are shown. 
As detailed below, the comparison between these barrier
values and the actual real market 
price lead us to the definition of a buy or sell action which will be shown to 
be able of exploiting trends present in the real data.

The empirical analysis considers both in-sample and out-of-sample
cases. The difference between the two is that for the former a unique
set of values 
\mbox{$(D_m,D_a,\alpha,\beta_m,\beta_a)$}
is used, calibrated with the whole 25-year dataset, while in the
latter case the parameters 
\mbox{$(D_m,D_a,\alpha,\beta_m,\beta_a)$}
are calibrated each
year, on the basis of the previous 15-year history. 
For both approaches we will use the cumulative distribution values $\mathit{Q}=0.05,0.1,0.25$
($5\%,~10\%,~25\%$, respectively).

\subsection{Conditioned trading signals}\label{S2.1.2}
To exploit the non-Markovian character of our
model, besides the opening value of the day, $s_0^{(l)}$, 
we can use the value of the first $t_p$ returns of the day, 
$r_1,\ldots,r_{t_p}$,\footnote{Again, for simplicity dependence on the day $l$ is 
understood.} 
to
condition the subsequent expected evolution of 
the index. 

In general, the conditioned probability of the aggregated returns 
$r_{t_p+1}+\cdots+r_{t_p+\tau}$ given the previous
ones $r_1,\ldots,r_{t_p}$ is obtained as the ratio of the joint
PDF's:
\begin{equation}
p(r,\tau|r_1,\ldots,r_{t_p})
=\dfrac{p(r,\tau;r_1,\ldots,r_{t_p})}{p(r_1,\ldots,r_{t_p})}.
\end{equation}
Using Eq. (\ref{eq_joint_explicit}), we obtain the explicit expression
\begin{eqnarray}
p(r,\tau|r_1,\ldots,r_{t_p})
&=&
\;\dfrac{\Gamma \left( \frac{\alpha+t_p+1}{2} \right) }{\pi^{\frac{1}{2}} 
\;\Gamma \left( \frac{\alpha+t_p}{2} \right)} 
\left(\sum_{t=t_p+1}^{t_p+\tau} a_t^2\,\beta_t^2 \right)^{-1/2}\cdot
\nonumber\\
&&
\cdot\;\dfrac{
\left[  1 + \dfrac{r^2}{\sum_{t=t_p+1}^{t_p+\tau} a_t^2\,\beta_t^2}+\left( \dfrac{r_1}{a_1\,\beta_1}\right)^2 
+\cdots+ \left( \dfrac{r_\tau}{a_\tau\,\beta_\tau} \right)^2 \right] ^{-\frac{\alpha +t_p+1}{2}}
}{
\left[  1 + \left( \dfrac{r_1}{a_1\,\beta_1}\right)^2 
+\cdots+ \left( \dfrac{r_{t_p}}{a_{t_p}\,\beta_{t_p}} \right)^2 \right] ^{-\frac{\alpha + t_p}{2}}
}.
\end{eqnarray}
In this way, with a simple change of variable the equation defining the conditioned quantile function,
\begin{equation}
\mathit{Q}=\int_{-\infty}^{r_{min,\tau}(\mathit{Q})} dr\;p(r,\tau|r_1,\ldots,r_{t_p}),
\end{equation} 
becomes
\begin{equation}
\label{eq_quantile_conditioned}
\mathit{Q}
=
\;\dfrac{\Gamma \left( \frac{\alpha+t_p+1}{2} \right) }{\pi^{\frac{1}{2}} 
\;\Gamma \left( \frac{\alpha+t_p}{2} \right)} 
\int_{-\infty}^{z_{min,\tau}(\mathit{Q})} dz\;[1+z^2]^{-\frac{\alpha +t_p+1}{2}},
\end{equation}
with
\begin{equation}
z_{min,\tau}(\mathit{Q})
\equiv
\dfrac{r_{min,\tau}(\mathit{Q})}{
\left(\sum_{t=t_p+1}^{t_p+\tau} a_t^2\,\beta_t^2 \right)^{1/2}
\left[  1 + \left( \dfrac{r_1}{a_1\,\beta_1}\right)^2 
+\cdots+ \left( \dfrac{r_{t_p}}{a_{t_p}\,\beta_{t_p}} \right)^2 \right] ^{\frac{1}{2}}
}.
\end{equation}
Given the previous returns $r_1,\ldots,r_{t_p}$, again Eq. (\ref{eq_quantile_conditioned})
can be solved numerically for $r_{min,\tau}(\mathit{Q})$.
Knowing $s_0^{(l)}$ the determination of the conditioned lower and upper barriers,
$s_{min,\tau}(\mathit{Q})$ 
and $s_{max,\tau}(\mathit{Q})$ respectively, proceeds then as indicated in the previous section.

\begin{figure}
\begin{center}
\resizebox*{!}{9cm}{\includegraphics{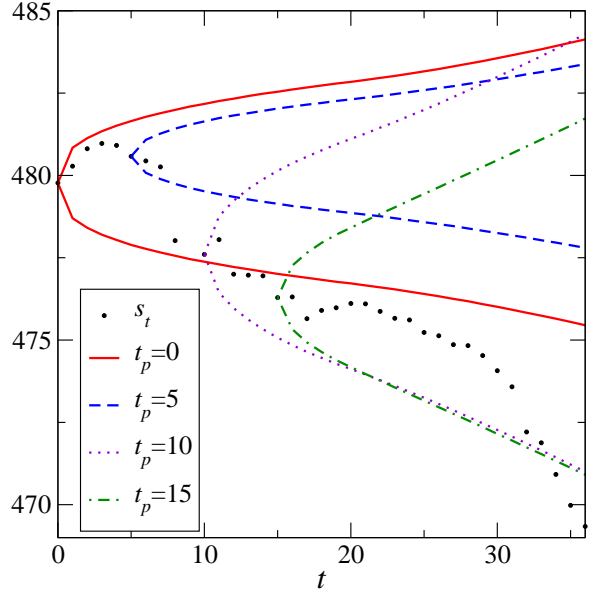}}
\caption{Upper and lower expected index values [$\mathit{Q}$=10\%] for one random
  day of the dataset (February 4th, 1994). As shown in the inset, different numbers of
  conditioning returns are considered.} 
\label{fig:day_2101}
\end{center}
\end{figure}

In Fig.~\ref{fig:day_2101} some interesting features of conditioned
trading signals can be detected. 
In particular, it is clear the influence of the
past returns to determine the expected amplitude of the next
ones.
Conditioned trading will thus be based on bounds which may vary
according to the information contained in the first returns of the day.

\section{Developing and applying intraday strategies}\label{S3}
When the index value $s_t^{(l)}$ breaks through the lower
$s_{min,\tau}(\mathit{Q})$ or the upper $s_{max,\tau}(\mathit{Q})$
values of the quantile function,
this violation can be used to define a trading strategy.
Note that the ensemble property we consider lasts from 10:00 
a.m. to 16:00 p.m. As a consequence, we define a trading strategy which
operates within this time lapse. 
To avoid the impact of
news arrivals in the close-to-open time frame, the
trading strategy opens and closes positions within the day. 
Furthermore, since we are using a $10$-minute dataset, density forecasts 
(and therefore the quantile functions at level $\mathit{Q}$ and $1-\mathit{Q}$) will be
available from $10:00\;{\rm a.m.}+(t_p+1)\times10\;{\rm min}$. 

Within a certain day $l$, given the specific values of the quantile function 
at level $\mathit{Q}$ and $1-\mathit{Q}$,
$s_{min,\tau}(\mathit{Q})$ and $s_{max,\tau}(\mathit{Q})$ respectively, the trading
signals and the trading activity are defined as follows: 
\begin{itemize}
\item[A:] If there are no open positions
\medskip
\begin{itemize}
\item[A.i:] Buy if $s_t^{(l)}>s_{max,t}(\mathit{Q})$ \& $s_{t-1}^{(l)}<s_{max,t-1}(\mathit{Q})$
\item[A.ii:] Sell if $s_t^{(l)}<s_{min,t}(\mathit{Q})$ \& $s_{t-1}^{(l)}>s_{min,t-1}(\mathit{Q})$
\end{itemize}
\medskip
\item[B:] If there are open positions
\medskip
\begin{itemize}
\item[B.i:] Close a long position if $s_t^{(l)}<s_{max,t}(\mathit{Q})$ \& $s_{t-1}^{(l)}>s_{max,t-1}(\mathit{Q})$
\item[B.ii:] Close a short position if $s_t^{(l)}>s_{min,t}(\mathit{Q})$ \& $s_{t-1}^{(l)}<s_{min,t-1}(\mathit{Q})$
\smallskip
\item[B.iii:] Close long or short positions if they are still in place at market closing.
\end{itemize}
\end{itemize}
By construction, multiple trades are possible within volatile
days. Differently, in trending days, single operations will take
place, while during stable days, no positions will be taken. 

The quantile functions built with our model assume the
absence of linear correlations and then, if positive linear correlations exist,
on average, an open trade will last more than expected and the
probability of a positive profit will be slightly greater
than the probability of a negative profit.
In the light of these considerations and of our model assumptions,
the choice of the $10$-minutes time interval can be now better explained. 
Indeed, it is a compromise between the need of a fine definition of the price trajectory
(for an efficient application of the strategy) and the need of having small enough
linear correlations notwithstanding the presence of non-linear correlations
(for a good accordance with the characteristics of the model).
Furthermore, we noticed that considering a one-minute time intervals results in poorer
performances of the trading strategy, likely because with such a short intervals
the trades have a shorter life and thus smaller average profit.

Our purpose is to monitor the performances of the trading strategy
defined above. Therefore, we simulate the time evolution of a trader using
our strategy and having an initial cash amount equal to 1
million. Given the previous remarks about the validity of the ensemble
properties and their link with the trading strategy, at the beginning
of the day and at market closing the simulated portfolio is entirely composed by
cash. Furthermore, in order to avoid losses larger than the portfolio
value when implementing short positions, we limit the investment value
to 90\% of the overall portfolio value\footnote{We do not take into
  account the margins generally required when creating short
  position. We motivate this choice by the need of evaluating the
  strategy abilities on both long and short trades without penalizing
  short positions, as would be the case when margins larger than 10\%
  would be required. Furthermore, given that the trades will last at
  maximum for 5 hours and 50 minutes (from 10:10 to 16:00 p.m.), we believe 
  that an implicit margin of 10\% will be sufficient.}. 
For symmetry,
we apply the same rule also for long positions. As a result, when
trades are created, 10\% of the portfolio remains in cash. Once a
signal is observed, the trade is executed at the price
$s_t^{(l)}$.

As mentioned above, three different quantile levels will be
considered: 
$5\%$, $10\%$, $25\%$, and we
simulate both unconditioned and conditioned quantiles for all these
three levels. 
Specifically, 
we calculate price barriers using the opening price only (unconditioned, $t_p=0$),
or conditioning also on the first 3, 6 or 9 returns ($t_p=3,6,9$, respectively).
Higher numbers of initial conditioning points are not used for
two main reasons: first, their introduction results in poorer
performances compared to those observed with up to 9 conditioning points;
second, the number of observations within the day is limited to 36 
and we prefer not to reduce too much the time frame
available for trading activity.

\begin{table}
\caption{Average profit per trade in basis points (1985-2010).}
\begin{center}					
\begin{tabular}{cccc}
\hline
 & 25\% & 10\% & 5\% \\
\hline
\multicolumn{4}{c}{All trades}\\
\hline
0 & 4.479 & 3.473 & 2.722 \\
3 & 4.862 & 4.501 & 3.716 \\
6 & 4.525 & 4.285 & 4.166 \\
9 & 4.200 & 4.608 & 3.862 \\
\hline
\multicolumn{4}{c}{Long trades}\\
\hline
0 & 4.519 & 3.551 & 4.587 \\
3 & 4.886 & 4.775 & 4.119 \\
6 & 4.430 & 4.391 & 4.561 \\
9 & 4.354 & 4.830 & 4.514 \\
\hline
\multicolumn{4}{c}{Short trades}\\
\hline
0 & 4.441 & 3.403 & 1.283 \\
3 & 4.838 & 4.234 & 3.347 \\
6 & 4.625 & 4.185 & 3.833 \\
9 & 4.039 & 4.395 & 3.280 \\
\hline
\end{tabular}
\end{center}
\begin{small}
The first column reports the conditioning elements (0 stands for no
conditioning). The first row shows the quantile level. 
\end{small}
\label{taba}
\end{table}

\subsection{In-sample results}
We first analyse the trading strategy in-sample, in order to evaluate
its abilities in terms of yearly profits and average return by trade
(in basis points). At this stage, we also verify the impact of the
different number of conditioning returns. 
Table (\ref{taba}) reports the average return of the trades generated
by the strategy during the period from October 1985 to October
2010. Profits are indicated as average basis points per trade, and
distinguishing also between long and short trades. This allows
to verify if the strategy better identifies signals in a specific
direction. 

Two elements clearly appear. 
The first one is that the profit
decreases with decreasing quantile values, as if stronger signals
provide smaller performances irrespective of the
trade sign. We explain such an unexpected result by the fact that
the linear correlations are so small that 
the effectiveness of the trading strategy is decreased when the
quantile barriers become steeper.

The second relevant comment refers to the relation between
average profit and conditioning information. We observe that 
conditioning the price barriers to the first returns of the day increases the
profit. This is
somewhat expected as during the first part of the day the model adapts
its behaviour to the most recent data, and it provides therefore a better
fit to the ensemble property within the remaining part of the
day. Again, such a result holds independently of the trade sign.
Regarding the number of conditioning values, we note that the relation
between conditioning points and profits is not monotonic, and has a
maximum between one and six conditioning points. 
As a consequence of this
result and of the previous one about the quantile/profit relation, 
the out-of-sample analyses discussed in the next section 
have been performed with three conditioning
points, which is the value with maximum profit for the 5\% and 25\% quantiles.
Table (\ref{taba}) reports just the average trade returns.
However, we also analysed the distribution and the moments of the trade
returns. The results (not reported and availbale upon request) show that
the distribution of trade returns is highly leptokurtic and right skewed
(asymmetry is positive and, on average, larger than 3).

Table (\ref{tabb}) reports the same quantities as Table (\ref{taba}),
but by year (we drop 1985 and 2010 where only part of the year was
available). In general, the use of a 25\% quantile provides the higher
average profit. Furthermore, we observe that the various trading strategies are concordant in showing negative values in the range 2004-2007, where the
market was clearly upward trending and in a low volatility phase. 
This is an expected outcome, since the model detects violations
which are associated with high price fluctuations. The largest profits are
located during 1986 and 1987, in a high volatility period. 
Overall, the average profit per trade is relatively small, and
occasionally even negative. 
On the other hand, during some specific market phases the trading strategy
provides large average profits per trade, see for instance the 90's. 

\begin{table}
\caption{Average profit per trade in basis points (yearly values).}
\begin{center}	
\begin{tabular}{cccccccccc}
\hline
Year &	\multicolumn{3}{c}{Unconditioned} &
\multicolumn{3}{c}{Conditioned 3 points} &
\multicolumn{3}{c}{Conditioned 6 points} \\	 
Quantile &	25\% &	10\% &	5\% & 25\% & 10\% &	5\% & 25\% & 10\% &	5\% \\
\hline
1986 & 11.41 & 11.14 &  7.88 & 12.04 & 13.13 & 11.85 & 11.84 & 10.96 & 11.19 \\
1987 & 13.37 & 15.75 & 14.94 & 14.99 & 18.97 & 16.35 & 13.79 & 18.68 & 19.57 \\
1988 &  8.94 & 10.02 &  8.74 & 11.16 & 12.44 & 15.50 &  8.73 & 12.46 & 11.55 \\
1989 &  7.21 &  7.33 &  4.82 &  7.33 &  7.77 &  5.85 &  9.53 &  8.75 &  7.33 \\
1990 &  8.78 &  5.42 &  3.33 &  9.73 &  6.40 &  5.20 &  9.68 &  6.39 &  4.82 \\
1991 &  7.45 &  5.39 &  8.26 &  8.64 &  8.92 &  6.28 &  9.16 & 10.73 & 10.10 \\
1992 &  4.36 &  1.97 & -1.72 &  4.50 &  5.38 &  4.34 &  5.59 &  4.68 &  5.99 \\
1993 &  3.11 & -0.05 & -0.49 &  3.07 &  3.76 &  1.01 &  2.20 &  2.98 &  3.99 \\
1994 &  4.06 &  2.90 & -2.52 &  4.57 &  3.83 &  1.35 &  4.12 &  4.49 &  4.07 \\
1995 &  3.11 & -0.01 &  1.73 &  3.05 &  2.58 &  1.92 &  2.67 &  3.08 &  0.65 \\
1996 &  2.28 & -0.22 & -4.56 &  3.07 & -0.50 & -4.21 &  3.91 &  3.54 &  1.75 \\
1997 &  4.3  &  1.83 & -1.06 &  5.41 &  3.99 &  2.83 &  7.04 &  5.36 &  2.20 \\
1998 &  6.15 &  6.42 &  5.39 &  5.90 &  6.43 &  2.76 &  4.90 &  3.30 &  3.05 \\
1999 &  2.27 &  1.39 & -2.70 &  3.07 &  3.29 &  1.11 &  2.58 &  1.28 &  0.00 \\
2000 &  3.33 &  0.91 &  1.87 &  2.53 &  4.08 &  2.94 &  2.50 &  2.33 &  1.09 \\
2001 &  1.04 &  1.25 &  0.65 &  1.40 &  0.91 &  2.95 &  2.32 &  0.30 & -0.49 \\
2002 &  7.50 &  6.94 &  7.94 &  5.30 &  3.14 &  5.73 &  3.97 &  0.93 &  1.79 \\
2003 &  2.04 &  1.56 &  1.59 &  3.64 &  2.77 &  1.29 &  2.18 &  1.90 &  2.76 \\
2004 &  1.27 &  2.98 &  3.08 &  1.01 &  1.48 &  2.20 &  1.40 &  3.21 &  2.97 \\
2005 &  1.85 &  2.18 &  1.15 &  2.47 &  1.08 & -1.00 &  3.13 &  2.43 &  4.26 \\
2006 &  0.75 & -1.85 & -5.16 &  0.56 &  0.48 & -1.32 &  0.53 &  0.47 & -0.14 \\
2007 & -0.49 & -2.07 & -2.40 &  0.96 &  1.60 &  1.12 &  0.48 &  1.49 & -1.18 \\
2008 &  8.24 &  3.70 &  4.08 &  7.18 &  2.49 &  0.01 &  5.64 &  2.26 &  2.34 \\
2009 &  1.82 &  1.87 &  0.49 &  3.91 &  0.62 & -0.01 &  0.89 & -1.57 & -0.24 \\
\hline
\end{tabular}
\end{center}
\label{tabb}
\end{table}

Besides the impact of quantile level and number of conditioning returns, a third element
is of interest: the number and type of trades created by the strategy
in a given time interval. Table (\ref{tabc}) reports several elements
for the most recent years. The first column just repeats the content
of Table (\ref{tabb}), while the following ones separately consider
Long and Short trades distinguishing between ``true'' and
``false'' signals. We call ``true'' a trading signal which really
provides a positive trade profit. As previously mentioned, 
the existence of ``false'' signals
is also influenced by the discreteness of our dataset
and the ratio true-to-false signals might not be optimal.

In light of this comment, the presence of substantial positive profits
despite the relatively small number of true
signals [see the last two columns of
Table (\ref{tabc})] should be considered as a positive, potentially
interesting outcome of the strategy. 
A further element supporting the strategy is the
average profit of trades originated by true signals. 
Both for long and short trades, these average profits are sensibly
larger, peaking at more than 150 basis points (note that trades are created within the day). 
On the contrary, 
false signals lead to trades with small losses (compared to the
gains), even if these losses are larger
during volatile market phases as in 2008 and 2009 
(both for long and short trades).
The number of trades is much larger when using 25\% quantiles compared
to 5\% quantiles; 
while there are small differences between the use of
Conditioned and Unconditioned quantiles. Long and short trades are
almost numerically equivalent both in 
trending and volatile market phases. 
As naturally expected, the number of trades 
increases during volatile periods,
irrespectively of the sign.
Finally, also the number of relatively few
true signals does not depend on the trade sign.
In summary, Table (\ref{tabc}) shows evidence of some potential
interest in the proposed strategy, since the average profit for true signals is
quite elevate (in particular compared to the overall average profit).

\begin{table}
\caption{Average profit per trade and number of trades: long/short trades, false/true signals.}
\begin{tabular}{cccccccccccccccccc}
\hline
 & \multicolumn{7}{c}{Average profit (bp)} & \multicolumn{9}{c}{Number of trades} \\
 & All & \multicolumn{3}{c}{Long} & \multicolumn{3}{c}{Short} & All & \multicolumn{3}{c}{Long} & \multicolumn{3}{c}{Short} & \multicolumn{2}{c}{\% True} \\
 & & All & True & False & All & True & False & & All & True & False & All & True & False & Long & Short \\
\hline
 & \multicolumn{16}{c}{25\% - Conditioned 3 points} \\
\hline
2005 & 2.47 &  2.93 &  38.72 &  -5.89 &  2.08 & 41.67 & -6.71 & 561 & 258 & 51 & 207 & 303 & 55 & 248 & 19.8\% & 18.2\% \\
2006 & 0.56 &  1.11 &  32.74 &  -5.50 &  0.07 & 36.63 & -6.55 & 566 & 266 & 46 & 220 & 300 & 46 & 254 & 17.3\% & 15.3\% \\
2007 & 0.96 & -1.30 &  39.94 &  -8.87 &  3.56 & 68.78 & -9.38 & 650 & 348 & 54 & 294 & 302 & 50 & 252 & 15.5\% & 16.6\% \\
2008 & 7.18 &  3.45 & 115.48 & -22.12 & 10.50 & 141.75 & -20.61 & 720 & 339 & 63 & 276 & 381 & 73 & 308 & 18.6\% & 19.2\% \\
2009 & 3.91 &  2.79 &  83.36 & -13.69 &  5.40 & 94.67 & -15.54 & 650 & 371 & 63 & 308 & 279 & 53 & 226 & 17.0\% & 19.0\% \\
\hline
 & \multicolumn{16}{c}{5\% - Conditioned 3 points} \\
\hline
2005 & -1.00 & 0.15 & 30.20 & -3.10 & -2.30 & 21.54 & -7.55 & 154 & 82 & 8 & 74 & 72 & 13 & 59 & 9.8\% & 18.1\% \\
2006 & -1.32 & 0.45 & 32.27 & -3.09 & -2.57 & 43.33 & -6.05 & 121 & 50 & 5 & 45 & 71 & 5 & 66 & 10.0\% & 7.0\% \\
2007 & 1.12 & 1.73 & 53.31 & -7.32 & 0.78 & 52.20 & -8.03 & 190 & 67 & 10 & 57 & 123 & 18 & 105 & 14.9\% & 14.6\% \\
2008 & 0.01 & -2.82 & 122.92 & -17.98 & 2.26 & 127.03 & -20.63 & 358 & 158 & 17 & 141 & 200 & 31 & 169 & 10.8\% & 15.5\% \\
2009 & -0.01 & 0.71 & 90.40 & -11.25 & -0.65 & 107.26 & -12.46 & 288 & 136 & 16 & 120 & 152 & 15 & 137 & 11.8\% & 9.9\% \\
\hline
 & \multicolumn{16}{c}{25\% - Unconditioned} \\
\hline
2005 & 1.85 & 1.53 & 41.88 & -6.65 & 2.12 & 42.9 & -6.27 & 542 & 249 & 42 & 207 & 293 & 50 & 243 & 16.9\% & 17.1\% \\
2006 & 0.75 & 0.05 & 34.58 & -6.31 & 1.42 & 42.82 & -7.38 & 489 & 238 & 37 & 201 & 251 & 44 & 207 & 15.5\% & 17.5\% \\
2007 & -0.49 & -2.19 & 39.94 & -9.33 & 1.5 & 78.14 & -11.57 & 654 & 352 & 51 & 301 & 302 & 44 & 258 & 14.5\% & 14.6\% \\
2008 & 8.24 & 5.1 & 127.17 & -28.45 & 11.38 & 156.4 & -26.54 & 771 & 385 & 83 & 302 & 386 & 80 & 306 & 21.6\% & 20.7\% \\
2009 & 1.82 & 2.86 & 92.27 & -19.27 & 0.66 & 97.78 & -20.65 & 717 & 378 & 75 & 303 & 339 & 61 & 278 & 19.8\% & 18.0\% \\
\hline
 & \multicolumn{16}{c}{5\% - Unconditioned} \\
\hline
2005 & 1.15 & 5.71 & 66.1 & -0.22 & -1.34 & 7.07 & -2.67 & 34 & 12 & 1 & 11 & 22 & 3 & 19 & 8.3\% & 13.6\% \\
2006 & -5.16 & -1.94 & 4.28 & -2.56 & -6.93 & 11.43 & -7.90 & 62 & 22 & 2 & 20 & 40 & 2 & 38 & 9.1\% & 5.0\% \\
2007 & -2.40 & -0.02 & 67.11 & -12.01 & -3.38 & 68.96 & -14.74 & 114 & 33 & 5 & 28 & 81 & 11 & 70 & 15.2\% & 13.6\% \\
2008 & 4.08 & 3.31 & 161.15 & -29.28 & 4.78 & 147.73 & -27.07 & 469 & 222 & 38 & 184 & 247 & 45 & 202 & 17.1\% & 18.2\% \\
2009 & 0.49 & 2.25 & 118.14 & -16.59 & -1.10 & 103.59 & -14.87 & 408 & 193 & 27 & 166 & 215 & 25 & 190 & 14.0\% & 11.6\% \\
\hline
\end{tabular}
\label{tabc}
\end{table}

While the previous tables where focusing on the average return over single
trades, Table (\ref{tabd}) focuses on the overall profit
of the strategy over single years (assuming a starting cash amount of
1 million). The returns are reported in percentages, and show evidence
of positive performances in most periods. Comparing first the
Conditioned versus Unconditioned quantiles, we observe that
conditioned modeling is clearly better: the associated returns are higher apart
from few cases,  and their standard deviation is in most cases smaller\footnote{The
  standard deviation is computed over the daily returns of the
  simulated portfolio and then annualized. Note that days without
  any trading signal provide zero returns, since we did not assume
  any remuneration for the bank account.}. 
Contrasting the 25\% and
5\% quantiles, the use of narrower bands (25\%) for the identification of the
signals provide larger returns over the years. This potentially
exposes the portfolio to a number of trades generated by false
signals, but the profits coming from true signals balance
them. Such a result holds irrespectively of the conditioning type. 
Finally, if we compare the performances of the trading strategy (25\%
quantiles) to that of the underlying equity index [see the last two
columns of Table (\ref{tabd})], we note a relevant positive result:
when the market is experiencing high volatility, our strategy provides
positive returns with a volatility smaller than that of the market,
and this is particularly evident when the market has yearly negative
returns; on the contrary, when the market is in a low volatility
period, our strategy has in some cases small or negative returns.
In general, the trading 
strategy has always a volatility smaller than that of the market. This
finding suggests that it could be used to hedge the market
volatility, since it provides positive returns in case of high market
volatility, and with smaller risk. This is further confirmed by the
correlation between market and our strategy returns and between market and
our strategy standard deviation [see the last row of Table
(\ref{tabd})]: positive and very high correlation in the case of standard
deviations, and low negative correlation for returns. Finally, if we compute the yearly Sharpe ratios (not reported) we can note that the strategies based on 25\% quantiles provide higher remuneration per unit of risk compared to the market.

\begin{table}
\caption{In-sample yearly return and standard deviation compared with the S\&P500 Index.}
\begin{tabular}{ccccccccccc}
\hline
 & \multicolumn{2}{c}{25\% - C. 3 p.} & \multicolumn{2}{c}{5\% - C. 3 p.} & \multicolumn{2}{c}{25\% - Unc.} & \multicolumn{2}{c}{5\% - Unc.} & \multicolumn{2}{c}{S\&P500} \\
 & Return & Dev.st & Return & Dev.st & Return & Dev.st & Return & Dev.st & Return & Dev.st \\
\hline
1986 & 70.15 &  7.91 & 23.00 &  5.60 & 63.61 &  7.76 &  6.66 &  4.02 &  14.62 & 14.64 \\
1987 & 89.63 & 15.18 & 38.62 & 11.32 & 79.65 & 18.03 & 23.56 & 16.22 &   2.03 & 32.01 \\
1988 & 64.62 &  9.48 & 29.69 &  6.52 & 49.62 &  9.94 & 10.58 &  6.69 &  12.40 & 17.02 \\
1989 & 38.48 &  7.59 &  8.55 &  5.90 & 32.48 &  7.26 &  2.81 &  5.73 &  27.25 & 13.01 \\
1990 & 60.70 &  8.19 & 14.07 &  5.19 & 51.33 &  8.38 &  5.66 &  4.70 &  -6.56 & 15.89 \\
1991 & 51.30 &  7.33 & 14.55 &  4.53 & 38.56 &  7.38 &  8.73 &  3.92 &  26.31 & 14.24 \\
1992 & 22.19 &  4.45 &  5.06 &  2.07 & 17.73 &  4.09 & -0.96 &  1.26 &   4.46 &  9.64 \\
1993 & 14.33 &  3.97 &  1.14 &  1.92 & 10.53 &  3.66 & -0.12 &  1.33 &   7.06 &  8.57 \\
1994 & 22.82 &  4.62 &  1.57 &  2.49 & 16.95 &  4.79 & -1.13 &  1.34 &  -1.54 &  9.80 \\
1995 & 14.93 &  3.68 &  1.76 &  2.10 & 12.24 &  3.90 &  0.49 &  1.54 &  34.11 &  7.78 \\
1996 & 16.91 &  5.36 & -5.55 &  2.19 & 12.17 &  5.77 & -3.46 &  2.78 &  20.26 & 11.73 \\
1997 & 35.58 &  9.32 &  5.81 &  6.15 & 29.74 &  9.88 & -2.50 &  6.15 &  31.01 & 18.06 \\
1998 & 38.57 &  9.56 &  6.20 &  4.78 & 40.12 & 11.47 & 11.98 &  7.42 &  26.67 & 20.21 \\
1999 & 20.60 &  7.66 &  2.89 &  4.03 & 14.68 &  9.79 & -6.62 &  4.93 &  19.53 & 18.00 \\
2000 & 16.06 & 10.46 &  6.88 &  6.18 & 24.10 & 13.02 &  5.62 &  8.80 & -10.14 & 22.13 \\
2001 &  9.15 & 10.32 &  6.20 &  5.26 &  6.30 & 11.96 &  2.36 &  7.04 & -13.04 & 21.47 \\
2002 & 39.53 & 12.82 & 15.27 &  8.99 & 61.15 & 16.42 & 33.80 & 12.86 & -23.37 & 25.93 \\
2003 & 24.70 &  7.92 &  2.27 &  3.84 & 12.71 &  9.53 &  3.82 &  5.11 &  26.38 & 17.00 \\
2004 &  5.64 &  6.06 &  3.26 &  2.81 &  6.42 &  6.09 &  2.13 &  1.71 &   8.99 & 11.05 \\
2005 & 13.81 &  4.58 & -1.41 &  1.85 &  9.57 &  4.55 &  0.40 &  1.42 &   3.00 & 10.24 \\
2006 &  2.98 &  4.41 & -1.45 &  1.88 &  3.12 &  4.77 & -2.89 &  0.91 &  13.62 &  9.99 \\
2007 &  4.61 &  7.46 &  1.86 &  3.60 & -4.31 &  7.92 & -2.59 &  3.57 &   3.53 & 15.93 \\
2008 & 53.19 & 19.62 & -2.24 & 13.19 & 68.93 & 23.97 & 14.81 & 20.24 & -38.49 & 40.81 \\
2009 & 25.01 & 12.56 & -0.46 &  7.24 & 11.18 & 15.49 &  1.16 & 10.84 &  24.71 & 27.18 \\
\hline
Corr. & -0.10 & 0.99 & -0.06 & 0.94 & -0.33 & 0.99 & -0.46 & 0.98 & & \\
\hline
\end{tabular}
\label{tabd}
\end{table}

\subsection{Out-of-sample results}
The above promising in-sample performances are confirmed in
the out-of-sample results. In this evaluation, we compare our
model to a more traditional approach, based on GARCH processes 
\cite{engle,bollerslev}. 
Shifting from the point of view of
ensembles to that of financial time series, several elements
characterizing high frequency data have to be considered. In particular, the
periodic behaviour of the intra-daily volatility has to be taken into
account \cite[][among others]{andersen_1,andersen_2}. 
To capture these
elements, together with variance asymmetry, we consider as a competing
model an asymmetric GARCH, the GJR \cite{glosten} with a
periodic deterministic variance component. Our choice is motivated by
the relative simplicity of the competitor, a kind of benchmark, and by
the possibility of easily generating from this model density forecasts
at a given quantile under a distributional assumption for the model
innovations. The competing model is given as follow: 
\begin{itemize}
\item the empirical returns on a 10-minute time scale are represented
  as: 
  $r_{t}^{(l)}=m_{t}^{(l)}\;\epsilon_{t}^{(l)}$,
  where $t$ identifies the 10-minute period within day $l$ with a
  range which is now $t=1,2,...T$ ($T=36$ for our dataset), $m_{t}^{(l)}$ is a deterministic 
  periodic function, and
  $\epsilon_{t}^{(l)}$ is the stochastic component; 
  this model
  implies that returns are generated as 
  $R_{t}^{(l)}=\mathbb N \left(0,m_{t}^{(l)}\;\mathbb{V}_{\epsilon}\right)$, 
  where
  $\mathbb N(\mu,\sigma)$ indicates a Gaussian random variable of mean $\mu$
  and variance $\sigma$, and 
  $\mathbb{V}_{\epsilon}$ is the (stochastic) variance of the random component; 
\item $m_{t}^{(l)}$ is a periodic deterministic variance modeled
  similarly to 
  \citet{andersen_2}, but using dummy variables instead of
  harmonics; we might represent returns as 
\begin{equation}
\label{FFF1}
\ln [(R_{t}^{(l)})^2]=
\ln[(m_{t}^{(l)})^2]+\ln[(\varepsilon_{t}^{(l)})^2],
\end{equation}
with 
\begin{equation}
\label{FFF2}
\ln[(m_{t}^{(l)})^2]=a_1 + \sum_{j=2}^{T} a_j d_{t,j}^{\,(l)},
\end{equation}
where $d_{t,j}^{\,(l)}$, $j=2,...T$ is a dummy variable assuming value 1
when $j=t$ and zero otherwise, 
while $a_1,a_2 \ldots a_{T}$  are
parameters to be estimated; 
\item furthermore, the stochastic term $\varepsilon_{t}^{(l)}$ follows a GJR
  model (Glosten et al. 1993) allowing thus for the
  decomposition 
\begin{equation}
\varepsilon_t^{(l)}=\sigma_t^{(l)}\;Z_{t}^{(l)},
\end{equation}
where $Z_{t}^{(l)}=\mathbb N(0,1)$ and the conditioned variance
is given by
\begin{equation}
(\sigma_{t}^{(l)})^2=\omega + \left(\alpha_0 + 
\alpha_1 \;I(\overline{\epsilon}_{t}^{\,(l)}<0) \right)
(\overline{\epsilon}_{t}^{\,(l)})^2 
+ \beta\;(\overline{\sigma}_{t}^{\,(l)})^2,
\end{equation}
where 
\begin{itemize}
\item $(\overline{\epsilon}_{t}^{\,(l)})^2\equiv(\epsilon_{t-1}^{(l)})^2$ if $t>1$ and
$(\overline{\epsilon}_{t}^{\,(l)})^2\equiv(\epsilon_{T}^{l-1})^2$ if $t=1$,
\item $(\overline{\sigma}_{t}^{\,(l)})^2\equiv(\sigma_{t-1}^{(l)})^2$ if $t>1$ and
$(\overline{\sigma}_{t}^{\,(l)})^2\equiv(\sigma_{T}^{l-1})^2$ if $t=1$,
\item $I(\overline{\epsilon}_{t}^{\,(l)}<0)$ is equal to one when 
  $\overline{\epsilon}_{t}^{\,(l)}$ is negative and zero otherwise,
\item $\omega$, $\alpha_0$, $\alpha_1$ and $\beta$ are parameters to
  be estimated. These parameters must satisfy the constraints for
  variance positivity and covariance stationarity $\omega>0$,
  $\alpha_0>0$, $\alpha_1>0$, $\beta>0$ and
  $\alpha_0+0.5\alpha_1+\beta<1$ (under an assumption of symmetry for
  the density characterizing $Z_{t}^{(l)}$. 
\end{itemize}
\end{itemize}
The estimation of the model proceeds by steps. At first the periodic
component is estimated by linear regression using equations
(\ref{FFF1}) and (\ref{FFF2}). The fitted periodic component is used
to recover the estimated values of $\epsilon_{t}^{(l)}$. Over those, the
GJR parameters are estimated by Quasi Maximum Likelihood approaches
using a Gaussian likelihood. Given the estimated parameters, and under
a Gaussian density for the innovations $z_{t}^{(l)}$, we generate possible
paths for the future evolution of the conditioned variance
$(\sigma_{t}^{(l)})^2$, of the innovations $\epsilon_{t}^{(l)}$, and of the
returns $r_{t}^{(l)}$ (the periodic component $m_{t}^{(l)}$ is purely
deterministic and is thus simply replicated in the forecasting
exercise). Under the distributional hypothesis, the needed quantiles
are then determined and used as an alternative input for the
identification of the trading signals. 
 
In Table (\ref{tabe}) we report the out-of-sample average profit per
trade, using the Unconditioned and Conditioned trading strategies
as well
as the one based on the GJR model. As mentioned earlier, the
out-of-sample evaluation focuses only in the range 2000 to 2010, since
the period 1985-1999 is used to calibrate the 
models. Results for our model are similar to in-sample outcomes,
with conditioned modeling providing better results. 
Both
Conditioned and Unconditioned model specifications have performances
largely better than the GJR model. The only case in which the
conditioned variance model have performances comparable to our
approach is over long trades and for quantiles equal to 25\% and 10\%.
Even in the out-of-sample case we analyse the trade returns
distribution, in particular contrasting our model's results
to the GJR returns. We note that the returns distribution from
the GJR model is characterised by smaller levels of both kurtosis
and skewness compared to our model's results. We note that also
the GJR returns distribution is right skewed (results are available upon requests).

The differences among the
strategies appear more clearly in Table (\ref{tabf}), which
contains annual returns of the simulated portfolios. We note here that
using the 25\% quantiles together with a conditioning on the first
three returns of the day provides the best results in high volatility
market phases (large than 20\% annualized daily market volatility). On
the contrary, when the volatility is lower, there is not a clear
preference across the models (GJR included). The yearly volatility of
the GJR is lower compared to our models as well as to the market, but
the yearly profits are clearly unsatisfactory. In addition, the last
row of Table (\ref{tabf}) 
reports the correlations between market and strategies returns and
between market and strategies standard deviations: results
confirm the previous in-sample findings for our strategies, while for the GJR we
have a lower correlation in the standard deviations and a positive correlation between the strategy returns and the market returns.

\begin{table}
\caption{Out-of-sample average profit per trade in basis points (2000-2010).}
\begin{center}
\begin{tabular}{cccc}
\hline
 & 25\%	& 10\% & 5\% \\
\hline
 & \multicolumn{3}{c}{All trades} \\			
\hline
Unc. & 2.865 & 2.249 & 1.992 \\
Cond. (3) & 2.830 & 1.758 & 1.383 \\
GJR & 0.286 & 0.397 & 0.569 \\
\hline
 & \multicolumn{3}{c}{Long trades} \\
\hline
Unc. & 2.226 & 2.601 & 3.921 \\
Cond. (3) & 2.444 & 1.514 & 1.870 \\
GJR & 2.512 & 2.576 & 0.855 \\
\hline
 & \multicolumn{3}{c}{Short trades} \\ 
\hline
Unc. & 3.488 & 1.950 & 0.459 \\
Cond. (3) & 3.203 & 1.990 & 0.965 \\
GJR & -1.313 & -1.404 & 0.320 \\
\hline
\end{tabular} 
\end{center}
\begin{small}
The first column reports the model: Unc. stands for our model with no
conditioning; Cond. (3) refers to the model with a conditioning to the
first three returns of the day; finally, GARCH identifies the
conditioned variance specification with deterministic periodic
component. The first rows reports the quantile level. 
\end{small}
\label{tabe}
\end{table}

\begin{table}
\caption{Out-of-sample yearly return and standard deviation compared with the S\&P500 Index.}
\begin{tabular}{cccccccccccccccc}
\hline
& \multicolumn{2}{c}{25\% - C. 3 p.} 
& \multicolumn{2}{c}{5\% - C. 3 p.} 
& \multicolumn{2}{c}{25\% - Unc.} 
& \multicolumn{2}{c}{5\% - Unc.} 
& \multicolumn{2}{c}{25\% - GJR} 
& \multicolumn{2}{c}{5\% - GJR} 
& \multicolumn{2}{c}{S\&P500} \\
& Return & Dev.st & Return & Dev.st & Return & Dev.st & Return & Dev.st & Return & Dev.st & Return & Dev.st & Return & Dev.st \\
\hline
2000 & 18.39 & 10.74 &  8.52 &  6.56 & 22.12 & 13.18 &  3.45 &  9.82 & -2.04 & 5.12 &  0.19 & 1.60 & -10.14 & 22.13 \\
2001 &  7.65 & 10.39 &  6.11 &  5.48 &  6.48 & 12.01 &  2.60 &  8.13 &  7.18 & 5.20 & -2.53 & 1.83 & -13.04 & 21.47 \\
2002 & 41.00 & 12.77 & 13.40 &  9.48 & 57.78 & 16.51 & 30.32 & 13.22 & -5.06 & 5.58 & -0.02 & 1.88 & -23.37 & 25.93 \\
2003 & 27.72 &  7.92 &  1.26 &  3.87 & 16.15 &  9.26 &  5.34 &  5.18 &  4.20 & 4.69 &  0.87 & 2.08 &  26.38 & 17.00 \\
2004 &  5.30 &  5.99 &  3.55 &  2.67 &  6.28 &  5.98 &  1.99 &  1.79 & -1.31 & 1.80 &  0.24 & 0.35 &   8.99 & 11.05 \\
2005 & 13.40 &  4.59 & -0.45 &  1.90 &  8.47 &  4.49 &  0.27 &  1.43 & -0.34 & 1.63 &  0.20 & 0.35 &   3.00 & 10.24 \\
2006 &  2.73 &  4.41 & -1.00 &  1.79 &  3.71 &  4.76 & -2.09 &  1.24 &  2.75 & 1.69 & -0.19 & 0.43 &  13.62 &  9.99 \\
2007 &  5.07 &  7.45 &  2.17 &  3.98 & -4.95 &  7.88 & -3.76 &  3.65 & -4.44 & 1.99 &  0.56 & 0.91 &   3.53 & 15.93 \\
2008 & 48.76 & 19.72 & -7.63 & 15.00 & 69.26 & 23.97 & 21.52 & 20.28 & -1.60 & 4.23 &  0.24 & 1.20 & -38.49 & 40.81 \\
2009 & 26.37 & 12.61 &  1.51 &  7.28 & 14.28 & 14.95 & -1.38 & 10.48 & 11.38 & 5.35 &  1.18 & 1.18 &  24.71 & 27.18 \\
\hline
Corr. & -0.50 & 0.99 & -0.07 & 0.98 & -0.73 & 0.99 & -0.72 & 0.99 & 0.52 & 0.68 & 0.41 & 0.49 &  &  \\
\hline
\end{tabular}
\label{tabf}
\end{table}

Overall, the results suggest the existence of some trading
opportunities with respect to the time series and sample size
used. Furthermore, they show how the combination of advanced
statistical methodologies could be used for the development of trading
rules or trading schemes which have a large resemblance with those
commonly used in technical analysis.

\section{Concluding remarks}
In this study we verified that one can model the daily, high frequency dynamics
of the S\&P index on the basis of the scaling 
properties of the aggregated returns. The resulting martingale
description generates histories whose statistical properties
are consistent with those of the ensemble of daily histories
on which the model has been calibrated.
In the version of the model developed in this work the scaling 
property has been generalized with respect to previous formulations
for FX rates returns \cite{bbcs}, to describe the
dynamics in a daily window of index evolution
encopassing the whole trading session, and with the average 
volatility varying in a non-monotonous way.
The martingale character of the model implies strictly zero
linear correlation between elementary returns. This condition is 
only approximately verified within the dataset. 
Empirical estimations indeed show that the linear correlation is small,
but nonzero, and changes sign as a function of time (see Fig. \ref{fig:lin_corr}).
On the other hand, we verified that our martingale model reproduces
very well the more substantial nonlinear correlations.

The presence of these linear correlations suggests that
the postulated daily process could present trends which,
althought difficult to model, may be exploited by appropriate 
trading strategies.
Our choice has been then to use the martingale forecast capability of our model to 
define a trend-following strategy which reveals the presence of trends in the data
in terms of a nonzero average profit.
Besides the potential applicative interest on which we comment below,
this result calls attention about the 
relevance that some apparently minor empirical features of the dynamics
may have. In our case such 
a feature is a not strict satisfaction of the martingale character
of the asset's dynamics, a property which is often assumed in theoretical modeling.   

Performing both in-sample and out-of-sample analyses
we demonstrated how the
scaling properties of the stochastic process can be used to
derive long-term (intraday) density forecasts. In turns,
these density forecasts can be used to
define trading signals and to implement an intra-day trading strategy
which exposes a small arbitrage opportunity. By
comparing the trading outcomes to those obtained from a standard
GARCH model, namely, the GJR \cite{glosten},
we showed better performances for the trading strategy based on the
proposed model. 
The average trade profit is limited over the entire time span, even though
local levels might be higher. 
Further studies aiming at improving the
trading strategy and the empirical application of the model are thus
required and under development. 

Summarizing our findings, we can say that the proposed model has
some potential for the development of trading strategies aimed at
hedging the volatility risk, since their performances are positive
during high market volatility, and characterized by a lower risk
compared to the market index. Signals extracted from the model 
could also be considered as confirmatory signals for
other strategies working with high frequency data, or could be used to
detect relevant market movements. 

In this empirical example we do not consider several elements that
could have an impact on the trading strategy profits. We motivate this
by the need of evaluating the model in comparison to a simple
benchmark. Across the elements we did not include, we have the trading
costs. Once those are introduced, the profits reported in the previous
tables would be sensibly reduced. However, the trading strategy we
implement is based on a fixed frequency database, using a 10-minute
interval. This has a relevant impact on the trading outcomes. In fact,
if a quantile violation is observed at time $t$, we execute the
trade with the price observed at this same point in
time. However, the violation could have taken place in any instant
in the ten minutes before $t$. A trader using our approach would
produce quantiles to be used for each period of 10 minutes, but would
immediately detect the violation, and operate in the market soon after
it (assuming she/he fully trusts the signal). On the contrary,
working with a fixed time span of 10 minutes, we lose part of the
potentially relevant content of the signal, since the price at time
$t$ might be significantly different from the real price observed
at the trade execution just after the violation occurred.

Another element not included in our trading example is the
remuneration of the bank account. In addition, overnight liquidity
operations could be introduced given that the portfolio is entirely
into cash from 16:00 p.m. of day $t$ up to 10:09 a.m. of day 
$t+1$. Finally, we note that even the trading strategy could be
improved, for instance introducing stop-loss and take-profit bounds on
the implemented orders.

\section*{Acknowledgments}
We thank M. Zamparo for useful discussions. 
This work is supported by 
``Fondazione Cassa di Risparmio di Padova e Rovigo'' within the 
2008-2009 ``Progetti di Eccellenza'' program.

\section*{Appendix: Testing linear and non-linear returns' correlations}
\begin{figure}
\begin{center}
\resizebox*{9cm}{!}{\includegraphics{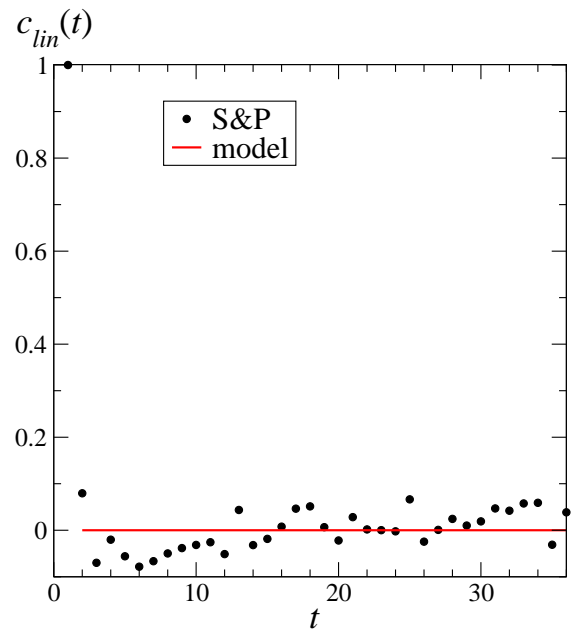}}
\caption{Linear correlation.}
\label{fig:lin_corr}
\end{center}
\end{figure}

First of all we show that linear correlations on a $10$-minute scale, 
\begin{equation}
c_{lin}(t) \equiv
\dfrac{\dfrac{1}{M}\sum_{l=1}^M\;r_1^{(l)}\;r_{t}^{(l)}}
{\sqrt{ \dfrac{1}{M} \sum_{l=1}^M \; \left(r_1^{(l)}\right)^2}
\;\sqrt{ \dfrac{1}{M}\sum_{l=1}^M\; \left(r_{t}^{(l)}\right)^2}},
\end{equation} 
albeit absent in the model oscillate around $\pm0.1$ in the empirical
data (see Fig.~\ref{fig:lin_corr}). 
These correlation are responsible for the presence of a trend
which makes the presented strategy profitable. 
Due to their oscillating nature, it is
difficult finding for them an appropriate  
modeling. 

\begin{figure}
\begin{center}
\resizebox*{8cm}{!}{\includegraphics{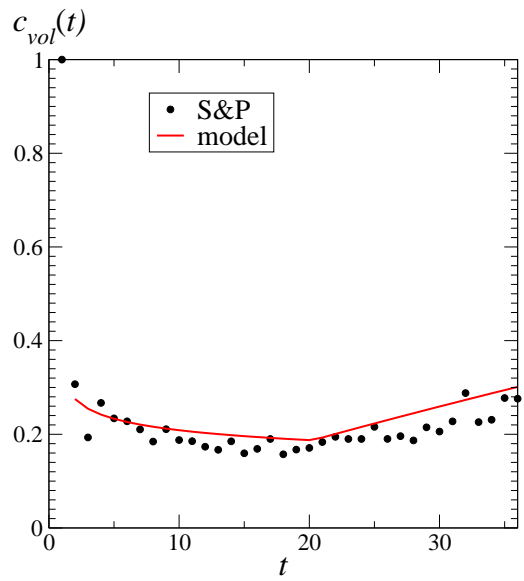}}
\caption{Volatility autocorrelation.}
\label{fig:volatility_autoc}
\end{center}
\end{figure}

However, non-linear correlations represented for instance by 
the volatility autocorrelation,   
\begin{equation}
c_{vol}(t) 
\equiv 
{
{
{1\over M}\sum_{l=1}^{M} |r_1^{(l)}| |r_t^{(l)}|
-\left({1\over M} \sum_{l=1}^{M} |r_1^{(l)}|\right) 
 \left({1\over M} \sum_{l=1}^{M} |r_t^{(l)}|\right)
}
\over
{{1\over M}\sum_{l=1}^{M} |r_1^{(l)}|^2 
-\left({1\over M} \sum_{l=1}^{M} |r_1^{(l)}|\right)^2}
},
\end{equation}
are a stronger and much more stable feature which is 
well reproduced by our model during both the morning  
and the afternoon trading sessions
(see Fig.~\ref{fig:volatility_autoc}, where 
the parameters derived from the in-sample analysis have been used). 

\label{lastpage}

\end{document}